\documentclass{birkjour_t2}

\usepackage[latin1]{inputenc}

\usepackage{amsmath}
\usepackage{amsfonts}
\usepackage{amssymb}
\usepackage{amsthm}

\usepackage{booktabs}
\usepackage{color}
\usepackage{enumerate}
\usepackage{graphicx}
\usepackage{hyperref}

\newcommand{\R}{\mathbb{R}}

\author{Z. Huang}
\email{rubyhuang87@gmail.com}
\address{University of Cambridge Computer Laboratory, Cambridge CB3 0FD, U.K.}

\author{M. England}
\email{Matthew.England@coventry.ac.uk}
\address{Coventry University Faculty of Engineering, Environment and Computing, Coventry, CV1 2JH, U.K.}

\author{D.J. Wilson}
\email{david.john.wilson@me.com}
\address{}

\author{J. Bridge}
\email{james.bridge@runbox.com}
\address{University of Cambridge Computer Laboratory, Cambridge CB3 0FD, U.K.}

\author{J.H. Davenport}
\email{J.H.Davenport@bath.ac.uk}
\address{Department of Computer Science, University of Bath, Bath, BA2 7AY, U.K.}

\author{L.C. Paulson}
\email{lp15@cam.ac.uk}
\address{University of Cambridge Computer Laboratory, Cambridge CB3 0FD, U.K.}

\title[Using Machine Learning to Improve Cylindrical Algebraic Decomposition]{Using Machine Learning to Improve \\ Cylindrical Algebraic Decomposition}

\begin{document}

\begin{abstract}

Cylindrical Algebraic Decomposition (CAD) is a key tool in computational algebraic geometry, best known as a procedure to enable Quantifier Elimination over real-closed fields.  However, it has a worst case complexity doubly exponential in the size of the input, which is often encountered in practice. It has been observed that for many problems a change in algorithm settings or problem formulation can cause huge differences in runtime costs, changing problem instances from intractable to easy.  A number of heuristics have been developed to help with such choices, but the complicated nature of the geometric relationships involved means these are imperfect and can sometimes make poor choices.   We investigate the use of machine learning (specifically support vector machines) to make such choices instead.  

Machine learning is the process of fitting a computer model to a complex function based on properties learned from measured data. In this paper we apply it in two case studies: the first to select between heuristics for choosing a CAD variable ordering; the second to identify when a CAD problem instance would benefit from Gr\"obner Basis preconditioning.  These appear to be the first such applications of machine learning to Symbolic Computation.  We demonstrate in both cases that the machine learned choice outperforms human developed heuristics.  

\end{abstract} 

\keywords{Symbolic Computation,  Computer Algebra, Machine Learning, Support Vector Machine, Cylindrical Algebraic Decomposition, Gr\"obner Basis, Parameter Selection. 
\hspace*{1.1cm} and 68T05 (Learning and Adaptive Systems)} 

\maketitle

\section{Introduction}

\subsection{Quantifier elimination}

A logical statement is \emph{quantified} if it involves the universal quantifier $\forall$ or the existential quantifier $\exists$.  The \emph{Quantifier Elimination} (QE) problem is to derive from a quantified formula an equivalent un-quantified one.  A simple example would be that the quantified statement, ``$\exists x$ such that $x^2 + b x + c = 0$'' is equivalent to the unquantified statement ``$b^2 - 4c \geq 0$'', when working over the real numbers.  

We can think of QE as simplifying or solving a problem.  The tools involved fall within the field of \emph{Symbolic Computation}, implemented in specialist \emph{Computer Algebra Systems}.  The exact nature of the tools mean we can draw insight from the answers:  the simple example of QE above showed how the number of solutions of the quadratic depends on the discriminant.  Further, many sentences which appear at first glance to not be quantified, in fact are. For example, the statement ``the natural logarithm is the inverse of the exponential function'' is really the statement ``$\forall x$ we have $\ln(e^x) = x$''.  Hence solutions to the QE problem have numerous applications in logic and computational mathematics, and in turn throughout engineering and the sciences.  Classical applications include motion planning in robotics \cite{WDEB13} while recent ones include artificial intelligence to pass a university entrance exam \cite{AMIA14}, the detection of Hopf bifurcations in chemical reaction networks \cite{EEGSSW15} and formal theorem proving \cite{Paulson2012}.

When the logical statements are expressed as \emph{Tarski formulae}, Bool\-ean combinations ($\land, \lor, \neg, \rightarrow$) of statements about the signs\footnote{principally $\{=0,>0,<0\}$, but the Boolean combinations allow $\{\ne0, \ge0, \le0\}$ as well} of polynomials with integer coefficients, then the QE problem is always soluble \cite{Tarski1948}.  However, the only implemented general QE procedure has algorithmic complexity doubly exponential in the number of variables \cite{DH88}, a theoretical result that is often experienced in practice.  For many problem classes QE procedures will work well at first, but as the problem size increases the doubly exponential wall is inevitably hit.  It is hence of critical importance to optimise the way QE procedures are used and the formulation of problems, to ``push the doubly exponential wall back'' and open up a wider range of tractable applications.

\subsection{Machine learning for QE}

A QE procedure can often be run in multiple ways to solve a problem: they can be initialized with different options (e.g. variable ordering \cite{DSS04}, equational constraint designation \cite{BDEW13}); tasks can be completed in different orders (e.g. order of constraint analysis \cite{EBCDMW14}); and the problem itself may be expressible in different formalisations (e.g. the different formulations of the ``piano mover's problem'' \cite{WDEB13}).  Changing these settings can have a substantial effect on the computational costs (both time and memory), the tractability, or even the theoretical complexity of a problem.  

Analysing these choices is a critical, and until recently understudied, problem.  At the moment most QE procedures either force their users to make these choices explicitly, or make them using fairly crude human-designed heuristics which have only a limited scientific basis.  
The present paper describes two case studies that use machine learning techniques to make these choices instead.  To the authors knowledge these are the first such applications of machine learning to QE, or Symbolic Computation more generally.  
One similar application has since followed by Kobayashi et al. \cite{KIMA16} who applied machine learning to decide the order of sub-formulae solving for their QE procedure\footnote{The feature set they used for their support vector machine was seeded by Table \ref{table:CSAFeature}.}.  Our work followed the application of machine learning by some of the authors to theorem proving \cite{HP13a, BHP14}.  

The two case studies both concern the Cylindrical Algebraic Decomposition (CAD) algorithm which takes a formula and produces a decomposition of real space from which a solution to a QE problem on the input can be constructed.  We consider the decision of variable ordering to use for CAD; and whether to precondition the input using Gr\"obner Basis technology.  Each decision can have a large effect on the time and memory requirements of the CAD construction.  We demonstrate that in each case the decision can be made using machine learning, more specifically a Support Vector Machine (SVM) with Radial Basis Function.

Machine learning, an over arching term for tools that allow computers to make decisions that are not explicitly programmed, usually involves the statistical analysis of large quantities of data.  On the surface this sounds at odds with the field of Symbolic Computation which prizes exact correctness rather than inexact numerics or probabilistic tools.  We are able to combine these fields because the choices we use machine learning to make are between different methods that would all produce correct and exact answers, but potentially with very different computational costs.   

\subsection{Paper outline}

We continue in Section \ref{SEC:Background} by introducing background material on CAD (Section \ref{SUBSEC:CAD}); machine learning (Section \ref{SUBSEC:ML}) and the topics of our two case studies (Sections \ref{SUBSEC:IntroCSA} and \ref{SUBSEC:IntroCSB}).  In Section \ref{SEC:Methodology} we describe the methodology of our work including the software employed (Section \ref{SUBSEC:Software}); the datasets used (Section \ref{SUBSEC:Data}) and how the data will be analysed and choices made (Section \ref{SUBSEC:DataEvaluation}).  We then give the detailed results of the two case studies in Sections \ref{SEC:CSA} and \ref{SEC:CSB} including the features used for the classification, the optimisation of the SVM parameters and the feature selection methods employed.  We finish in Section \ref{SEC:Conclusion} with a summary and possibilities for future work.

The two case studies have been presented in conference proceedings \cite{HEWDPB14} and \cite{HEDP16}.  This journal article draws them together into a single work, expanding on various details, in particular: the preceding motivation, the discussion on suitable datasets in Section \ref{SUBSEC:Data} and the possibilities for future work in Section \ref{SUBSEC:Future}.

\section{Background Material}
\label{SEC:Background}

\subsection{Cylindrical algebraic decomposition}
\label{SUBSEC:CAD}

A \emph{Cylindrical Algebraic Decomposition} (CAD) is a \emph{decomposition} of ordered $\mathbb{R}^n$ space into cells arranged \emph{cylindrically}: meaning the projections of any pair of cells with respect to the given ordering are either equal or disjoint.  The word algebraic is actually short for \emph{semi-algebraic} meaning that each CAD cell can be described with a finite sequence of polynomial constraints.  

A CAD is produced to be invariant for input, most commonly: \emph{sign-invariant} for a set of polynomials (so each polynomial is either positive, zero or negative throughout each cell); or \emph{truth-invariant} for formulae (so each formula is either true of false throughout each cell).

CADs and the first algorithm to compute them were introduced by Collins in 1975 \cite{Collins1975}.  A traditional CAD performs two stages.  First, \emph{projection} calculates sets of projection polynomials $S_i$ in variables $(x_1, \dots, x_i)$ by applying an operator recursively to the input to derive corresponding problems in lower dimensions.  Then in the \emph{lifting} stage CADs are built incrementally by dimension according to these polynomials.  First, the real line is decomposed according to the roots of the univariate polynomials $S_1$. Then over each cell $c$ in that decomposition, the bivariate polynomials $S_2$ are taken at a sample point and a decomposition of $c \times \R$ is produced according to their roots.  The theory of the projection operators allows working at a sample point to be sufficient to draw conclusions for the entire cell.  Taking the union of these then gives the decomposition of $\R^2$ and we proceed this way to a decomposition of $\R^n$.  For further details on the original CAD algorithm see the 1984 work of Arnon, Collins and McCallum \cite{ACM84I}.  

CAD has worst case complexity doubly exponential in the number of variables \cite{DH88} applicable due to the nature of the data structure rather than the method of computation \cite{BD07}.  For some applications there do exist algorithms with better complexity \cite{BPR06}, but CAD implementations still remain the best general purpose approach for many.  This is due to the extensive research to improve the efficiency of CAD since Collins' original work.  Prominent advances include: 
\begin{itemize}
\item improvements to the projection operator \cite{Hong1990, McCallum1998, Brown2001a, HDX14, MH16}; 
\item only constructing cells when necessary (e.g. partial CAD \cite{CH91} and sub-CAD \cite{WBDE14}); 
\item symbolic-numeric lifting schemes \cite{Strzebonski2006, IYAY09}.  
\end{itemize}
Despite it now being over 40 years since Collins' original algorithm CAD is still an active area of research with recent advances including:
\begin{itemize}
\item making use of any Boolean structure in the input \cite{BDEMW13, BDEMW16, EBD15}; 
\item decompositions via complex space \cite{CMXY09, CM14b, BCDEMW14}; 
\item local projection approaches \cite{BK15, Strzebonski2016}; 
\item non-uniform CADs (which relax the global cylindricity condition) \cite{Brown2015}; and their interaction with (Satisfiability Module Theory) SMT-solvers \cite{JdM12, AAB+16a}.
\end{itemize}

Via its use in QE, CAD has many application ranging from epidemic modelling \cite{BENW06} to formal verification \cite{PQ08}.  CAD also has independent applications: such as reasoning with multi-valued functions \cite{DBEW12} (where we identify the branch cuts of the single valued functions implemented in computer algebra systems \cite{EBDW13}, decompose space according to these, and check the validity of proposed simplifications from the multi-valued theory); and generating semi-algebraic descriptions of regions in parameter space where multi-stationarity can occur for biological networks \cite{BDEEGGHKRSW17, EEGRSW17}.  
New applications are being found all the time, such as structural design (minimizing the weight of trusses) \cite{CC18}; the derivation of optimal numerical schemes \cite{EH16}; and the validation of economic hypotheses \cite{Mulligan2016}.

\subsection{Machine learning}
\label{SUBSEC:ML}

Machine learning deals with the design of programs that can learn rules from data.  This is an attractive alternative to manually constructing such rules when the underlying functional relationship is very complex, as appears to the be the case for the decisions to be made in our two case studies.  Machine learning techniques have been widely used in a great many fields, such as web searching \cite{BFJ96}, text categorization \cite{Sebastiani2002}, robotics \cite{SV00}, expert systems \cite{FR86}.

Machine learning has a long history.  We reference only a few relevant highlights here and refer the reader to, for example, the textbook of Alpaydin \cite{Alpaydin2004} for more details.  Seminal contributions include: the first computational model for \emph{neural networks} of McCulloch and Pitts \cite{MP43}; the \emph{perceptron} proposed by Rosenblatt as an iterative algorithm for supervised classification of an input into one of several possible non-binary outputs \cite{Rosenblatt1958}; \emph{decision tree learning} approaches  \cite{PK14}, which apply serial classifications to refine the output state; the \emph{multi-layer perceptron} \cite{HSW89} (a modification that can distinguish data that are non-linearly separable), and the original \emph{support vector machine} algorithm \cite{VC64}, a development of the perception approach.  

\emph{Support Vector Machines} (SVMs) are the machine learning tool used for our case studies.  
The standard SVM classifier takes a set of input data and predicts one of two possible classes from the input, making them a non-probabilistic binary classifier. An SVM model represents the examples as points in space, mapped so that the points of the categories are divided by a clear gap, as wide as possible.  The original work \cite{VC64} used linear classification, but SVMs can also efficiently perform a non-linear classification by mapping their inputs into high-dimensional feature spaces using a kernel function \cite{BGV92, CV95}. 

SVMs today give a powerful and robust method for both classification and regression. 
\emph{Classification} refers to the assignment of input examples into a given set of classes (the output being the class labels).  \emph{Regression} refers to a supervised pattern analysis in which the output is real-valued.  Given a set of examples, each marked as belonging to one of the two classes, an SVM training algorithm builds a model that assigns new examples into one of the classes. 
Modern SVM technology can deal efficiently with high-dimensional data, and is flexible in modelling diverse sources of data. 

As noted above, a modern SVM will typically use a kernel function to map data into a high dimensional feature space. Kernel functions enable operations in feature space without ever computing the coordinates of the data in that space. Instead they simply compute the inner products between all pairs of data vectors. This operation is generally computationally cheaper than the explicit computation of the coordinates. We refer the reader to the textbook of Shawe-Taylor and Cristianini \cite{SC04} for further details on kernel functions.

The success of SVMs has led to a rapid spread of the use of machine learning.  For example the survey paper of Byun and Lee \cite{BL03} identifies their use in face recognition,  object detection, handwriting recognition, image retreieval and speech recognition; while the textbook of Sch{\"o}lkopf, Tsuda and Vert details applications in computational biology \cite{STV04}.  This paper describes two case studies which are the first applications (to the best of the authors' knowledge) in computer algebra software.

\subsection{Case Study A topic $-$ CAD variable ordering}
\label{SUBSEC:IntroCSA}

A CAD is defined with respect to a variable ordering (the cylindricity condition is that the projections of any two cells onto a smaller coordinate space \emph{with respect to the ordering} is either equal or disjoint).  The variable ordering dictates the order in which variables are eliminated during the projection phase and the order of subspaces in which CADs are produced during lifting.  For example, a variable order of $z \succ y \succ x$ would mean that cells in the CAD of $(x, y, z)$-space should have equal or disjoint projections when projected onto $(x, y)$-space or $x$-space.  A CAD algorithm would eliminate $z$ first and then $y$ during projection; and during lifting would first decompose the $x$-axis and then the $(x, y)$-plane.

Depending on the application requirements the variable ordering may be determined, constrained, or entirely free (as with the application to branch cut analysis of multi-valued functions described above).  The most common application, QE, requires that the variables be eliminated in the order in which they are quantified in the formula but makes no requirement on the free variables.  For example, we could eliminate the quantifier in 
\[
\exists x \quad ax^2 + b x + c = 0
\]
using any CAD which eliminates $x$ first.  The other variables could be studied in any order and  so there are six possibilities we can choose from when running the CAD algorithm.    Our own CAD implementation builds a CAD for the polynomial under ordering $a \prec b \prec c$ with only 27 cells, but uses 115 for the reverse ordering.
Also, note that since we can switch the order of quantified variables in a statement when the quantifier is the same, we also have some choice on the ordering of quantified variables.  For example, a QE problem of the form $\exists x \exists y \forall a \, \phi(x, y, a)$ could be solved by a CAD under either ordering $x \succ y \succ a$ or ordering $y \succ x \succ a$.

It has long been documented that this choice can dramatically affect the feasibility of a problem.  In fact, Brown and Davenport presented a class of problems in which one variable ordering gave output of double exponential complexity in the number of variables and another output of a constant size \cite{BD07}.  Heuristics have been developed to help with this choice, with Dolzmann~et~al.~\cite{DSS04} giving the best known study.  After analysing a variety of metrics they proposed a polynomial degree based heuristic (the heuristic sotd defined below).  However, in CICM 2013 some of the present authors demonstrated examples for which that heuristic could be misled \cite{BDEW13}.  We provided an alternative (the heuristic ndrr below) which addressed the intricacies of those examples, but we saw for other examples that ndrr had its own shortcomings.  

Our thesis became that the best heuristic to use is dependent upon the problem considered.  However, the relationship between the problems and heuristics is far from obvious.  Hence we investigate whether machine learning can help with these choices, both for CAD itself and QE by CAD.  

We identified the following three heuristics for picking a CAD variable ordering in the literature:

\begin{description}
\item[Brown] This heuristic chooses a variable ordering according to the following criteria, starting with the first and breaking ties with successive ones:
\begin{enumerate}[(1)]
\item Eliminate a variable first if it has lower overall degree in the input.
\item Eliminate a variable first if it has lower (maximum) total degree of those terms in the input in which it occurs.
\item Eliminate a variable first if there is a smaller number of terms in the input which contain the variable.
\end{enumerate}
It is labelled after Brown who documented it in the notes to an ISSAC tutorial \cite{Brown2004}.

\item[sotd] This heuristic constructs the full set of projection polynomials for each permitted ordering and selects the ordering whose corresponding set has the lowest sum of total degrees for each of the monomials in each of the polynomials. It is labelled sotd for \emph{sum of total degree} and was the outcome of the study by Dolzmann, Seidl, and Sturm which found it to be a good heuristic for both CAD and QE by CAD \cite{DSS04}.

\item[ndrr] This heuristic constructs the full set of projection polynomials for each ordering and selects the ordering whose set has the lowest number of distinct real roots of the univariate polynomials within.  It is labelled ndrr for \emph{number of distinct real roots} and was shown by Bradford et al. to assist with examples where sotd failed \cite{BDEW13}.
\end{description}  

Brown's heuristic has the advantage of being very cheap, since it acts only on the input and checks only simple properties. The ndrr heuristic is the most expensive (requiring real root isolation), but is the only one to explicitly consider the real geometry of the problem, rather than the geometry in complex space measured by the degrees.  

All three heuristics may identify more than one variable ordering as a suitable choice.  In this case we take the heuristic's choice to be the first of these after they had been ordered lexicographically.  This final choice will depend on the convention used for displaying the variable ordering.  For example, the software \textsc{Qepcad} and the notes where Brown introduces his heuristic \cite{Brown2004} use the convention of ordering variables from left to right so that the last one is projected first; while the algorithms in \textsc{Maple} and the papers introducing sotd and ndrr \cite{DSS04, BDEW13} use the opposite convention.  For this case study the heuristics were implemented in \textsc{Maple} and so ties were broken by picking the first lexicographically on the second convention.  This corresponds to picking the first under a reverse lexicographical order under the \textsc{Qepcad} convention.  The important point here is that all three heuristics had ties broken under the same convention and so were treated fairly.

\subsection{Case Study B topic $-$ Preconditioning CAD with a Gr\"obner basis}
\label{SUBSEC:IntroCSB}

A \emph{Gr\"obner Basis} $G$ is a particular generating set of an ideal $I$ defined with respect to a monomial ordering.  One definition is that the ideal generated by the leading terms of $I$ is generated by the leading terms of $G$.  Gr\"obner Bases (GB\footnote{We use the abbreviation GB for both Gr\"obner Basis and Gr\"obner Bases as grammar requires.}) allow for the calculation of properties of the ideal such as dimension and number of zeros.  They are one of the main practical tools for working with polynomial systems.  The GB definition, their properties and the first algorithm to derive one was introduced by Buchberger in his PhD thesis of 1965 (republished in English as \cite{Buchberger2006}).

Like CAD, there has been much research to improve GB calculation, with the $F_5$ algorithm \cite{Faugere2002} probably the most used approach today.  The calculation of a GB is, also like CAD, necessarily doubly exponential in the worst case \cite{MM82} (at least when using a lexicographic monomial ordering).  Despite this, the computation of GB can often be done very quickly and is usually a superior tool to CAD for a problem involving only polynomial equalities.  

From this arises the natural question: is the process of replacing a conjunction of polynomial equalities in a CAD problem by their GB a useful precondition for CAD?  I.e. let $E = \{e_1, e_2, \dots\}$ be a set of polynomials; 
$G = \{g_1, g_2, \dots\}$ be a GB for $E$; and $B$ be any Boolean combination of constraints, $f_i \, \sigma_i \, 0$, where $\sigma_i \in \{ <, >, \leq, \geq, \neq, =\}$) and $F = \{f_1, f_2, \dots\}$ is another set of polynomials.  Then the two  formulae
\begin{align*}
\Phi &= (e_1 = 0 \land e_2 = 0 \land \dots) \land B \mbox{ and } \\
\Psi &= (g_1 = 0 \land g_2 = 0 \land \dots) \land B
\end{align*}
are equivalent and a CAD truth-invariant for either could be used to solve problems involving  $\Phi$ (such as eliminating any quantifiers applied to $\Phi$).  So is it worth producing $G$ before calculating the CAD?

The first attempt to answer this question was given by Buchberger and Hong in 1991 \cite{BH91} who experimented with then recent implementation of GB \cite{BGK85} and  CAD \cite{CH91} (both in \textsc{C} on top of the \textsc{SAC-2} system \cite{Collins1985}).  Of the ten test problems they studied: 6 were improved by the GB preconditioning, with the speed-up varying from 2-fold to 1700-fold; 1 problem resulted in a 10-fold slow-down; 1 timed out when GB preconditioning was applied, while it would complete without it; and the other 2 were intractable both for CAD alone and the GB preconditioning step.  

The problem was revisited 20 years later by Wilson et al. \cite{WBD12_GB}.  The authors recreated the experiments of Buchberger and Hong \cite{BH91} using \textsc{Qepcad-B} \cite{Brown2003b} for the CAD and \textsc{Maple-16} for the GB.  As we would expect, there had been a big decrease in the computation timings, especially the GB: the two test problems previously intractable \cite{BH91} could now have their GB calculated quickly.  However, two of the CAD problems were still hindered by GB preconditioning.
The experiments were extended to: a wider example set (an additional 12 problems); the alternative CAD implementation in \textsc{Maple-16} \cite{CMXY09}; and the case where we further precondition by reducing inequalities of the system (the set $F$ above) with respect to the GB.  The key conclusion remained that GB preconditioning would in general benefit CAD (sometimes significantly) but could on occasion hinder it (to the point of making a tractable CAD problem intractable).

The authors of \cite{WBD12_GB} also defined a metric to assist with the decision of when to precondition, the \emph{Total Number of Indeterminates} (\texttt{TNoI}) of a set of polynomials $A$:
\begin{equation}
\label{eq:TNoI}
\texttt{TNoI}(A) = \textstyle \sum_{a \in A} \texttt{NoI}(a)
\end{equation}
where $\texttt{NoI}(a)$ is the number of indeterminates\footnote{In the context of the present paper indeterminates is synonymous with variables.  For the underlying application there may be a distinction between variables and parameters of the problem: but CAD, and correspondingly these metrics, would treat both the same.} in a polynomial $a$.  Then their heuristic was to build a CAD for the preconditioned polynomials only if the \texttt{TNoI} decreased following preconditioning.  For most test problems in the study the heuristic made the correct choice, but there were examples to the contrary, and little correlation between the change in \texttt{TNoI} and level of speed-up / slow-down.

For Case Study B we consider if machine learning can be applied to the decision of whether preconditioning CAD input with GB is beneficial for a particular problem.  We work on the reasonable assumption that GB computation is cheap for the problems on which CAD is tractable (in fact CAD must compute resultants which overestimate the GB \cite{ED16a}).  Hence we allow the classifier to use algebraic features of both the input problem and the GB itself to decide \emph{whether we want to use} the GB.

\section{Methodology}
\label{SEC:Methodology}

\subsection{Software}
\label{SUBSEC:Software}

\subsubsection{Computer algebra software}

For each case study we focussed on a single CAD implementation.  

Case Study A used \textsc{Qepcad} \cite{Brown2003b}: an interactive command line program written in C whose name stands for \textbf{Q}uantifier \textbf{E}limination with \textbf{P}artial \textbf{CAD}.  It is a competitive implementation of both CAD and QE that allows the user a good deal of control and information during its execution.  We used \textsc{Qepcad} with its default settings which implement McCallum's projection operator \cite{McCallum1998} and partial CAD \cite{CH91}.  It can also makes use of an equational constraint automatically (via the projection operator \cite{McCallum1999b}) when one is explicit in the formula, (where \emph{explicit} means the formula is a conjunction of the equational constraint with a sub-formula). 

Case Study B used the CAD algorithm in \textsc{Maple-17} that implements the Regular Chains based algorithm \cite{CMXY09}.  This is part of the \texttt{RegularChains} Library\footnote{\url{http://www.regularchains.org}} \cite{CM14d}, \cite{CM16} whose CAD procedures differ from the traditional projection and lifting framework of Collins.  They instead first decompose $\mathbb{C}^n$ cylindrically and then refine to a CAD of $\mathbb{R}^n$.  Previous experiments showed this implementation has the same issues of GB preconditioning as the traditional approach \cite{WBD12_GB}.  

All other computer algebra computations were done using tools in \textsc{Maple} and took minimal computation time:
\begin{itemize}
\item The variable ordering heuristics for Case Study A were calculated using tools in the authors \texttt{ProjectionCAD} library \cite{EWBD14}.  
\item The GB calculations for Case Study B used the implementation that ships with \textsc{Maple}: a meta algorithm calling multiple GB implementations.  The GBs were computed with a purely lexicographical ordering of monomials based on the same variable ordering as the CAD. 
\item The calculation of the algebraic features for both machine learning experiments was done using a mixture of build in \textsc{Maple} commands and the \texttt{ProjectionCAD} library.
\end{itemize}

\subsubsection{Machine learning software}
\label{SUBSUBSEC:SoftwareML}

Both case studies use the package \textsc{SVM-Light}\footnote{\url{http://svmlight.joachims.org}} \cite{Joachims1999} for the machine learning computations.  This software is an implementation of SVMs in C which  consists of two programs: \textsc{SVM learn} which fits the model parameters based on the training data and user inputs (such as the kernel function and the parameter values); and \textsc{SVM classify} which uses the generated model to classify new samples.

\textsc{SVM-Light} calculates a hyperplane of the $n$-dimensional transformed feature space, which is an affine subspace of dimension $n - 1$ dividing the space into two, corresponding to the two distinct classes.  \textsc{SVM-Classify} outputs margin values which are a measure of how far the sample is from this separating hyperplane. Hence the margins are a measure of the confidence in a correct prediction. A large margin represents high confidence in a correct prediction. The accuracy of the generated model is largely dependent on the selection of the kernel functions and parameter values.

\subsection{Datasets}
\label{SUBSEC:Data}

\subsubsection{Historic issues}

Despite its long history and significant software contributions the Computer Algebra community had a lack of substantial datasets in use, a significant barrier to machine learning.

There have been attempts to address this problem, such as the 1990s projects PoSSo and FRISCO for polynomials systems and symbolic-numeric problems respectively.  PoSSO collected a series of benchmark examples for GB, and a descendant of these can still be found online\footnote{\url{http://www-sop.inria.fr/saga/POL/}}.  However, this does not appear to be maintained and the polynomials are not stored in a machine-readable form.  Polynomials from this list still turn up in various papers, but there is no systematic reference, and it is not clear whether people are really referring to the same example (several of the examples are families which means that a benchmark has to contain specific instances).  The PoSSo Project aimed for ``level playing field'' comparisons, but at the time different implementations ran on different hardware / operating systems meaning this was not really achievable. 

The topic of benchmarking in computer algebra has most recently been taken up by the SymbolicData Project\footnote{\url{http://wiki.symbolicdata.org}} \cite{GNJ14} which is beginning to build a database of examples in XML format (although currently not with any suitable for CAD).  The software described in \cite{HL15} was built to translate problems in that database into executable code for various computer algebra systems.  The authors of \cite{HL15} discuss the peculiarities of computer algebra that make benchmarking particularly difficult including the fact that results of computations need not be unique and that the evaluation of the correctness of an output may not be trivial (or may be the subject of research itself). 

For CAD there are a number of ``classic'' example that reappear throughout the literature, such as those from \cite{BH91} and \cite {DSS04}.  Some of the present authors made an effort to collect these as a freely available online resource\footnote{\url{http://researchdata.bath.ac.uk/69/}} \cite{WBD12_EX} which contained references to the literature and encodings for various computer algebra systems.

\subsubsection{Dataset for Case Study A}

The resource discussed above \cite{WBD12_EX} did not have a sufficient number of CAD problems to run a machine learning experiment.  
Hence we decided to use instead the \texttt{nlsat} dataset\footnote{\url{http://cs.nyu.edu/~dejan/nonlinear/}} produced to evaluate the work in \cite{JdM12}.  The main sources of these examples are: \textsc{MetiTarski} \cite{Paulson2012}, an automatic theorem prover for theorems involving real-valued special functions (it applies real polynomial bounds and then using real QE tools like CAD); problems originating from
attempts to prove termination of term-rewrite systems; verification conditions from Keymaera \cite{PQR09}; and parametrized generalizations of the problems from \cite{Hong1991}.  

The problems from the \texttt{nlsat} dataset are all fully existential (the only quantifier is $\exists$) making them satisfiability or SAT problems.  They could thus be tackled by the new generation of SMT-solvers \cite{JdM12}.  Note that our decision to use only SAT problems was based solely on availability of data rather than it being a requirement of our technology, and the conclusions drawn are likely to be applicable outside of the SAT context.  

We extracted 7001 three-variable CAD problems for the experiment.  The number of variables was restricted for two reasons. First to make it feasible to test all possible variable orderings and second to avoid the possibility that \textsc{Qepcad} will produce errors or warnings related to well-orientedness with the McCallum projection \cite{McCallum1998}.  

Two experiments were undertaken, applying machine learning to CAD itself and to QE by CAD.  The problems for the former were obtained by simply removing all quantifiers.  We performed separate experiments since for quantified problems \textsc{Qepcad} can use the partial CAD techniques to stop the lifting process early if the outcome is already determined, while the full process is completed for unquantified ones and the two outputs can be quite different.

\subsubsection{Dataset for Case Study B}

For Case Study B we are more restricted as we need problems that are expressed with a conjunction of at least two equalities in order for their to be a non-trivial GB.  From the \texttt{nlsat} dataset 493 three-variable problems and 403 four-variable problems fit this criteria, which should have been a sufficient quantity for a machine learning experiment.  GB preconditioning was applied to each problem and cell counts from computing the CAD with the original polynomials and their replacement with the GB were computed and compared.  For every one of these problems the GB preconditioning was beneficial or made no difference; surprising as the experiments on much smaller datasets \cite{BH91, WBD12_GB} had shown much greater volatility.  It seems that a great deal of the problems involved had no solution to just the equational part (a fact the GB quickly identified) and so no CAD was required.  The \texttt{nlsat} dataset will need to be widened if it is to be used more extensively for computer algebra research, something that is now underway as part of the \textsf{SC}$^2$ Project\footnote{\url{http://www.sc-square.org/}} \cite{AAB+16a}

Since existing datasets were not suitable for the experiment, we had no choice but to  generate our own problems.  The generation process aimed for an unbiased dataset which would be computationally feasible for computing multiple CADs, and have some comparable structure (number of terms and polynomials) to existing CAD problems. 

We generated 1200 problems using the random polynomial generator \texttt{randpoly} in \textsc{Maple-17}.  Each problem has two sets of three polynomials; the first to represent conjoined equalities and the second for the other polynomial constraints (respectively $E$ and $F$ from the description in Section \ref{SUBSEC:IntroCSB}).  The number of variables was at most $3$, labelled $x, y, z$ and under ordering $x \prec y \prec z$; the number of terms per polynomial at most 2; the coefficients were restricted to integers in $[-20, 20]$; and the total degree was varied between 2, 3 and 4 (with 400 problems generated for each).  
A time limit of 300 CPU seconds was set for each CAD computation (all GB computations completed quickly) from which 1062 problems finished to constitute the final dataset.  Of these, 75\% benefited from GB preconditioning.   

We acknowledge the restrictions of random examples, and that the degree bound will result in a higher likelihood of the equations forming zero dimensional ideas.  But we note that our randomly generated dataset matches the results of \cite{BH91, WBD12_GB} with respect to most problems benefiting from GB, but a substantial minority not.  So we consider it suitable for the experiment.

\subsection{Evaluating the data}
\label{SUBSEC:DataEvaluation}

\subsubsection{Evaluating CADs}

All computations were performed on a 2.4GHz Intel processor, however this is not actually relevant as we evaluated the CAD performance using cell counts instead of timings.  We note that numerous previous studies have shown the CAD cell count to be closely correlated to timings.  Using cell count as the measure has the advantages of being discrete, machine independent and (up the theory used) implementation independent. It also correlates with the cost of any post-processing of the CAD (as would be required to produce an equivalent quantifier free formula for example).  The time taken to run the heuristics in Case Study A and to compute the GB in Case Study B, were minimal compared to the CAD computations and so not considered in the experiments.

For Case Study A, since each problem has three-variables and all the quantifiers are the same, all six possible variable orderings are admissible.  
In the quantified case \textsc{Qepcad} can collapse stacks when sufficient truth values for the constituent cells have been discovered to determine a truth value for the base cell.  Since our problems are all fully existential, the output for all quantified problems is therefore simply either true or false.  So it was not the number of cells in the output that was used to make decisions but instead the number of cells constructed during the process.  The best ordering was defined as the possibility resulting in the smallest cell count, (and if more than one ordering gives the minimal both orderings are considered the best).

For Case Study B the variable ordering was assumed fixed so only two CADs needed to be produced for a given example: one each for the input both before and after GB preconditioning.  The preconditioning was deemed to be beneficial if the CAD produced following it had less cells that it would have otherwise.  

\subsubsection{Making a choice}  
\label{SUBSUBSEC:Choice}

For Case Study A each problem has six possible orderings to choose from.  Rather than choosing between them directly we used SVMs to choose between the choice of the three previously identified heuristics (which seemed to excel on different problems).  This is both easier for the current problems and could be applied to those with far more ordering possibilities.  

Three classifiers were trained to predict whether each heuristic would make the correct choice for a given problem.  In a simple scenario only one of the three classifiers would return a positive result for any given problem (making the choice of best heuristic for the problem immediate).  However, in reality more than one classifier will return a positive result for some problems, while no classifiers may return a positive for others. Thus, instead we used the relative magnitudes of the classifiers in our experiment. The classifier with most positive (or least negative) margin value was selected to indicate the recommended decision procedure for the problem (we note that all features are  normalised to avoid scaling issues).  

For Case Study B the decision of whether to precondition was already a binary choice and there was only one previously known heuristic.  So for this experiment we decided to include the metric of that heuristic (TNoI) as a feature of the SVM, aiming to improve upon it with the inclusion of other features to produce the machine learned choice.  Thus only a single classifier was trained with its decision being the machine learned choice.

\section{Case Study A: CAD Variable Ordering}
\label{SEC:CSA}

In this section we give the full details of the machine learning experiment for Case Study A, starting with the problem features identified in Section \ref{SUBSEC:CSAFeatures}, then the parameter optimisation that we performed for the classifier in Section \ref{SUBSEC:CSAOptimization}, and the results in Section \ref{SUBSEC:CSAResults}.  We finish by comparing the existing heuristics on the entire problem set in Section \ref{SUBSEC:ISSACPoster} (a detour from the machine learning experiment but prompted due to the interesting results of Section \ref{SUBSEC:CSAResults}).

\subsection{Problem features}
\label{SUBSEC:CSAFeatures}

For both experiments (quantified and quantifier free), the datasets were randomly split into training sets (3545 problems each), validation sets (1735 problems each) and test sets (1721 problems each).

To apply machine learning, we need to identify features of the CAD problems that might be relevant to the correct choice of the heuristics. A feature is an aspect or measure of the problem that may be expressed numerically. Table \ref{table:CSAFeature} lists the 11 features that we identified.  Here $(x_{0}, x_{1}, x_{2})$ are the three variable labels that are used in all of our problems.  The proportion of a variable occurring in polynomials is the number of polynomials containing the variable divided by total number of polynomials. The proportion of a variable occurring in monomials is the number of terms containing the variable divided by total number of terms in all polynomials.

The number of features is quite modest, compared to other machine learning experiments.  They were chosen as easily computable features that describe the algebraic structure which could affect the performances of the heuristics.  

\begin{table}[h]
  \caption{The features used to classify examples in Case Study A}
  \label{table:CSAFeature}
  \setlength{\tabcolsep}{10pt}
  \def\arraystretch{1.2}%
  \centering
    \begin{tabular}{c l} 
      \hline
      Feature Number & Description \\  \hline \hline
      1 & Number of polynomials. \\ 
      2 & Maximum total degree of polynomials. \\ 
      3 & Maximum degree of $x_{0}$ among all polynomials. \\
      4 & Maximum degree of $x_{1}$ among all polynomials. \\
      5 & Maximum degree of $x_{2}$ among all polynomials. \\ 
      6 & Proportion of $x_{0}$ occurring in polynomials. \\
      7 & Proportion of $x_{1}$ occurring in polynomials. \\
      8 & Proportion of $x_{2}$ occurring in polynomials. \\ 
      9 & Proportion of $x_{0}$ occurring in monomials. \\ 
      10 & Proportion of $x_{1}$ occurring in monomials. \\
      11 & Proportion of $x_{2}$ occurring in monomials. \\ \hline
    \end{tabular}
\end{table}

Each problem contributes a feature vector of the training set which is then associated with a label: $+1$ (positive examples) or $-1$ (negative examples), indicating in which of two classes it was placed.  To take Brown's heuristic as an example, a corresponding training set was derived with each problem labelled $+1$ if Brown's heuristic suggested a variable ordering with the lowest number of cells, or $-1$ otherwise. 

The features could all be easily calculated from the problem input using \textsc{Maple}.  For example. if the input formula is defined using the set of polynomials
\[
\{
-6x_{0}^2-x_{2}^3-1,  \quad
x_{0}^4x_{2}+9x_{1},  \quad
x_{0}+x_{0}^2-x_{2}x_{0}-5
\}
\]
then the problem will have the feature vector 
\[
\left[ 
3, 5, 4, 1, 3, 1, \frac{1}{3}, 1, \frac{5}{9}, \frac{1}{9}, \frac{1}{3} 
\right].
\]
After the feature generation process, the training data (feature vectors) were  normalized so that each feature had zero mean and unit variance across the set.  The same normalization was then also applied to the validation and test sets.

\subsection{Parameter optimisation}
\label{SUBSEC:CSAOptimization}

As discussed in Section \ref{SUBSEC:ML}, SVMs use kernel functions to map the data into higher dimensional spaces where the data may be more easily separated.  The software we use, \textsc{SVM-Light}, has four standard kernel functions: linear, polynomial, sigmoid tanh, and radial basis function \cite{SC04}. For each kernel function, there are associated parameters which must be set. An earlier experiment by some of the authors applying machine learning to an automated theorem prover \cite{Bridge2010} compared these four kernels and found the Radial Basis Function (RBF) kernel performed the best in finding a relation between the simple algebraic features and the best heuristic choice.  The similarity of that context to the present one meant that we selected the same kernel was selected for this experiment.

The RBF function is defined as
\begin{equation}
\label{eq:RBF}
K(x, x\prime ) = \exp \left( -\gamma ||x - x\prime ||^2 \right)
\end{equation}
where $K$ is the kernel function, $x$ and $x\prime$ are feature vectors. There is a single parameter $\gamma$ in the RBF kernel function (\ref{eq:RBF}): as $\gamma$ rises the SVM seeks a closer fit, but runs the risk over-fitting \cite{SC04}.
Besides $\gamma$, two other parameters are involved in the SVM fitting process. The parameter $C$ governs the trade-off between margin and training error, while the cost factor $j$ is used to correct imbalance in the training set. Given a training set, we can easily compute the value of parameter $j$ by looking at the sign of the samples and setting $j$ equal to the ratio between negative and positive samples. However, it is not so trivial to find the optimal values of $\gamma$ and $C$. 

In machine learning, \emph{Matthew's correlation coefficient} (MCC) \cite{Matthews1975, BBCAN00} is often used to evaluate the performance of the binary classifications.  It is defined as 
\begin{equation}
\label{eq:MCC}
{\rm MCC} = \frac{{\rm TP}*{\rm TN}-{\rm FP}*{\rm FN}}{\sqrt{({\rm TP}+{\rm FP})({\rm TP}+{\rm FN})({\rm TN}+{\rm FP})({\rm TN}+{\rm FN})}},
\end{equation}
where: TP is the number of true positives, TN is the number of true negatives, FP is the number of false positives and FN is the number of false negatives. The denominator is set to $1$ if any sum term is zero. This measure has the value $1$ if perfect prediction is attained, $0$ if the classifier is performing as a random classifier, and $-1$ if the classifier exactly disagrees with the data. 

A grid-search optimisation procedure was used with the training and validation set, involving a search over a range of $(\gamma, C)$ values to find the pair which would maximize MCC in (\ref{eq:MCC}). We tested a commonly used range of values in our grid search process \cite{HCL03}: $\gamma$ took values from $\{2^{-15}, 2^{-14}, 2^{-13}, \dots, 2^{3}\}$; and $C$ took values from $\{2^{-5}, 2^{-4},$ $2^{-3}, \dots, 2^{15}\}$. Following the completion of the grid-search, the values for the parameters giving optimal MCC results were selected for each of the three CAD heuristic classifiers.  We also performed a similar calculation, selecting parameters to maximise the $F_1$-score \cite{Joachims2005}, but the results using MCC were superior.

The classifiers with optimal $(\gamma, C)$ were applied to the test set to output the margin values described in Section \ref{SUBSUBSEC:SoftwareML}.  As discussed in Section \ref{SUBSUBSEC:Choice} the machine learned choice is taken as the heuristic with most positive (or least negative) margin value.

\subsection{Results}
\label{SUBSEC:CSAResults}

We use the number of problems for which a selected variable ordering is optimal to measure the efficacy of each heuristic separately, and of the heuristic selected by machine learning.   

Table \ref{table:CSAsubset} breaks down the results into a set of mutually exclusive outcomes that describe all possibilities.  The column headed Machine Learning indicates the heuristic selected by the machine learned model with the next three columns indicating each of the fixed heuristics tested.  For each of these four heuristics, we may ask the question ``Did this heuristic select the optimal variable ordering?''  A Y in the table indicates yes and an N indicates no, with each of the 13 cases listed covering all possibilities.  Note that at least one of the fixed heuristics must have a Y since, by definition, the optimal ordering is obtained by at least one heuristic, while if they all have a Y it is not possible for machine learning to fail.  For each of these cases we list the number of problems for which this case occurred for both the quantifier free and quantified experiments.

\begin{table}
  \caption{Categorising the problems into a set of mutually exclusive cases characterised by which heuristics were successful.}
  \label{table:CSAsubset}
\centering
  \setlength{\tabcolsep}{6pt}
  \def\arraystretch{1.2}%
    \begin{tabular}{c c c c c c c} 
      \hline 
      Case & Machine Learning & sotd & ndrr & Brown & Quantifier Free & Quantified \\  \hline \hline
      1  & Y  & Y & Y & Y & 399 & 573 \\ \hline
      2  & Y  & Y & Y & N & 146 & 96 \\
      3  & N  & Y & Y & N & 39  & 24  \\ \hline 
      4  & Y  & Y & N & Y & 208  & 232  \\
      5  & N  & Y & N & Y & 35   & 43  \\ \hline 
      6  & Y  & N & Y & Y & 64  & 57  \\
      7  & N  & N & Y & Y & 7  & 11   \\ \hline 
      8  & Y  & Y & N & N & 106  & 66  \\
      9  & N  & Y & N & N & 106  & 75  \\ \hline 
      10 & Y  & N & Y & N & 159 & 101  \\
      11 & N  & N & Y & N & 58  & 89  \\ \hline 
      12 & Y  & N & N & Y & 230 & 208 \\
      13 & N  & N & N & Y & 164 & 146 \\ \hline 
    \end{tabular}
\end{table}

For many problems more than one heuristic selects the optimal variable ordering and the probability of a randomly selected heuristic giving the optimal ordering depends on how many pick it.  For example, a random selection would be successful $1/3$ of the time if one heuristic gives the optimal ordering, or $2/3$ of the time if two heuristics do so. 

In Table \ref{table:CSAsubset}, case $1$ is where machine learning cannot make any difference as all heuristics are equally optimal.  We compare the remaining cases in pairs (as indicated in the table). Each pair is constructed by keeping the behaviour of the fixed heuristics identical and looking at the two cases where machine learning picked a winning heuristic (one of the ones with a Y) or not. We see in every pair that machine learning succeeds far more often than it fails for.  For each pair we can compare with a random heuristic selection. For example, consider the pair formed by cases 2 and 3: where sotd and ndrr are successful heuristics and Brown is not.  A random selection would be successful $2/3$ of the time. For the quantifier free examples, machine learned selection is successful $146/(146+39)$ or approximately 79\% of the time, which is significantly better. 

We repeated this calculation for the quantified case and the other pairs, as shown in Table \ref{table:CSAright}.  In each case the values have been compared to the chance of success when picking a random heuristic, and so there are two distinct sets in Table \ref{table:CSAright}: those where only one heuristic was optimal and those where two are.   We see that machine learning performed stronger for some classes of problems than others.  For example, in quantifier free examples when only one heuristic is optimal machine learning does considerably better if that one is ndrr; while if only one is not optimal machine learning makes a poorer choice on average if that one is Brown.  Nevertheless, the machine learning selection is better than random in every case in both experiments.

\begin{table}
  \caption{Proportion of examples where machine learning picks a successful heuristic.}
\centering
  \label{table:CSAright}
  \setlength{\tabcolsep}{10pt}
  \def\arraystretch{1.2}%
    \begin{tabular}{c c c c c} 
      \hline 
      sotd & ndrr & Brown & Quantifier Free & Quantified \\  
      \hline \hline
       Y & Y & N & 79\% ($>$67\%) & 80\% ($>$67\%)  \\  
       Y & N & Y & 86\% ($>$67\%) & 84\% ($>$67\%)  \\  
       N & Y & Y & 90\% ($>$67\%) & 84\% ($>$67\%) \\  \hline
       Y & N & N & 50\% ($>$33\%) & 47\% ($>$33\%)  \\  
       N & Y & N & 73\% ($>$33\%) & 53\% ($>$33\%)  \\  
       N & N & Y & 58\% ($>$33\%) & 59\% ($>$33\%)  \\ \hline 
    \end{tabular}
\end{table}

By summing the numbers in Table \ref{table:CSAsubset} in which Y appears in a row for the machine learned selection and each individual heuristic, we get Table \ref{table:qf}. This compares, for both the quantifier free and quantified problem sets, the learned selection with each of the existing heuristics on their own. 

\begin{table}
  \caption{Total number of problems for which each heuristic picks the best ordering.}
  \label{table:qf}
  \centering
  \setlength{\tabcolsep}{10pt}
  \def\arraystretch{1.2}%
    \begin{tabular}{lcccc}
      \hline
                            & Machine Learning	& sotd	& ndrr & Brown	\\
      \hline \hline
      Quantifier free       & 1312	            & 1039	& 872  & 1107	\\
      Quantified            & 1333	            & 1109	& 951  & 1270	\\
      \hline
    \end{tabular}
\end{table}

For the quantifier free problems there were 399 problems where every heuristic picked the optimal, 
499 where two did and 
823 where one did.  Hence for this problem set the chance of picking a successful heuristic at random is 
\[
\frac{100}{1721} \left( 399 + 499*\tfrac{2}{3} + 823*\tfrac{1}{3} \right) \simeq 58\%
\]
which compares with $100 * 1312/1721 \simeq 76\%$ for machine learning.
For the quantified problems the figures are 
$64\%$ for a random choice 
and $77\%$ for a machine learned choice.   
Hence machine learning performs significantly better than a random choice in both cases.  

We can also compare to using only the heuristic that performed the best on this data, the Brown heuristic.  With that we would pick a successful ordering for
$64\%$ of the quantifier free problems and 
$74\%$ of the quantified problems.  So we see that a machine learned choice is also superior to using any one heuristic, with the improvement most impressive for the quantifier free problems.

\subsection{Further analysis of the existing heuristics}
\label{SUBSEC:ISSACPoster}

Of the three existing heuristics, Brown seems to be the best, albeit by a small margin.  Its performance was a little surprising to us, both because the Brown heuristic is not so well known (having never been formally published) and because it requires little computation (taking only simple measurements on the input).  We hence took a detour from our machine learning experiment to compare the performance of the three existing heuristics on the full problem set.

We first calculated in Table \ref{tab:issac1} for how many problems (and thus for what percentage) of the entire problem set each heuristic made the most competitive selection.  We see again Brown's heuristic is most likely to make the best choice, both when quantified and when quantifier free.

\begin{table}[t]
\caption{Comparison of how often each of the existing heuristics makes the best selection on the entire problem set}
  \label{tab:issac1}
  \centering
  \setlength{\tabcolsep}{10pt}
  \def\arraystretch{1.2}%
\begin{tabular}{lcccc}
\hline
                      & sotd			& ndrr				& Brown	  			\\
\hline \hline
Quantifier free       & 4221 (60.29\%)	& 3620 (51.71\%)  	& 4523 (64.61\%)	\\
Quantified            & 4603 (65.75\%)	& 4000 (57.13\%)  	& 5166 (73.79\%)    \\
\hline
\end{tabular}
\end{table}

\begin{figure}
\caption{Box plots of the percentage cell count savings made by the existing heuristics}
\label{fig:issac}
\begin{center}
\includegraphics[width=0.7\textwidth]{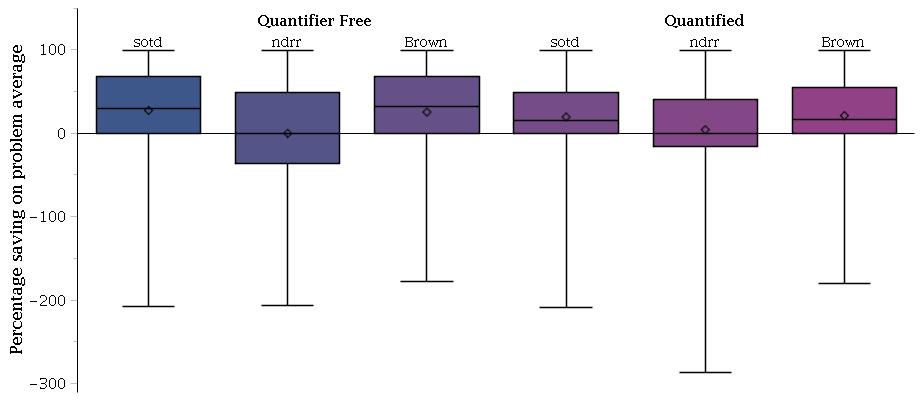}
\end{center}
\end{figure}

We next investigate how much of a cell count saving is offered by each heuristic.
We made the following calculations for each problem:
\begin{enumerate}[(1)]
\item The average cell count of the six orderings.
\item The difference between the cell count for each heuristic's pick and the problem average.
\item The value of (2) as a percentage of (1).
\end{enumerate}
These calculations were made for all problems in which no chosen variable ordering timed out (5262 of the quantifier free problems and 5332 of the quantified problems).  The data is visualised in the box plots of Figure \ref{fig:issac}, where the boxes indicate the second and third quartiles.  The mean and median values are given in Table \ref{tab:issac2} (and marked on the plots with circles and lines respectively). 

\begin{table}[t]
\caption{Average percentage cell count savings of the existing heuristics}
\label{tab:issac2}
\centering
\begin{tabular}{lcccccc}
\hline
                      & \multicolumn{3}{c}{Mean average} & \multicolumn{3}{c}{Median value} \\
\cmidrule(lr){2-4}
\cmidrule(lr){5-7}
                      &	sotd 	& ndrr 		& Brown	  &	sotd 	& ndrr 	 & Brown    \\
\hline \hline
Quantifier free       & 27.32\% & -0.20\% 	& 25.28\% & 29.47\% & 0.00\% & 32.28\%	\\
Quantified            & 19.47\% & 4.15\%   	& 21.03\% & 14.68\% & 0.00\% & 16.67\%  \\
\hline
\end{tabular}
\end{table}

While Brown's heuristic makes the best choice the most frequently, for the quantifier free problems the average saving of using sotd is actually higher. Ndrr performs the worst on average, but as we saw in the previous section there is a substantial class of problems where it makes a better choice than the others. This is illustrated if we consider the problems excluded from the previous discussion: those where at least one ordering timed out.  Table \ref{tab:issac3} describes how often each heuristic avoids a time out.  We see that for quantified problems ndrr does the best.

\begin{table}[t]
\caption{How often each heuristics avoids a time out on the set of problems where this occurred for at least one choice}
\label{tab:issac3}
\centering
\begin{tabular}{lccc}
\hline
                      & sotd	& ndrr  & Brown \\
\hline \hline 
Quantifier free       & 559		& 537   & 594	\\
Quantified            & 512     & 530   & 478   \\
\hline
\end{tabular}
\end{table}

\section{Case Study B: Preconditioning CAD with a Gr\"obner Basis}
\label{SEC:CSB}

In this section we give the full details of the machine learning experiment for Case Study B, starting with the problem features identified in Section \ref{SUBSEC:CSBFeatures}, then the initial experiment and results obtained in Section \ref{SUBSEC:CSBInitial}, the feature selection process that was run leading to improved results in Section \ref{SUBSEC:CSBSelection}, and finally a comparison with the human developed heuristic in Section \ref{SUBSEC:TNoI}.

\subsection{Problem features}
\label{SUBSEC:CSBFeatures}

Table \ref{table:CSBAllFeatures} shows the 28 problem features we identified for Case Study B.  They fall into three sets: those generated from the polynomials in the original problem ($\#1-12$), those obtained after applying GB preconditioning ($\#13-25$), and those involving both ($\#26-28$).  There is one more feature for the input after GB because the number of polynomials before GB was the same for each problem.

In the feature descripions $x$, $y$, and $z$ are the three variable labels used in the problems (the variable ordering was fixed as $x \prec y \prec z$).  The abbreviations \texttt{tds} and \texttt{stds} stand for \emph{maximum total degree} and \emph{sum of total degrees} respectively:
\begin{align*}
\texttt{tds}(F) &= \max_{f \in F} \; \texttt{tds}(f),
\qquad 
\texttt{stds}(F) = \sum_{f \in F} \texttt{tds}(f).
\end{align*}
Note that the \texttt{stds} measure differs from the \texttt{sotd} heuristic from Section \ref{SUBSEC:IntroCSA}. This is because \texttt{stds} measures the input polynomials only, while \texttt{sotd} measures the full set of CAD projection polynomials, and so is much more expensive.

We started with those features used for Case Study A and extended by including the simpler degree measure and the metric \texttt{TNoI} (see equation (\ref{eq:TNoI}) from Section \ref{SUBSEC:IntroCSB}).  Finally we included the base 2 logarithm of the ratio of differences between some of the key metrics.  All features could be calculated immediately with \textsc{Maple}.

In addition to training a classifier using all the features in Table \ref{table:CSBAllFeatures}, we trained classifiers using two subsets: one containing the features labelled $1-12$ concerning the original set of polynomials; and one with the features labelled $13-25$ concerning the polynomials after GB preconditioning.  We refer to the first subset as \emph{before features}, the second as \emph{after features} and the full set as \emph{all features}.

\begin{table}[t]
\label{table:CSBAllFeatures}  
\caption{Initial feature set to classify examples in Case Study B}
  \centering
    \begin{tabular}{c l} 
      \hline
      Feature Number  & Description \\  \hline \hline
      1 & \texttt{TNoI} before GB. \\ 
      2 & \texttt{stds} before GB. \\ 
      3 & \texttt{tds} of polynomials before GB.\\
      4 & Max degree of $x$ in polynomials before GB. \\
      5 & Max degree of $y$ in polynomials before GB. \\ 
      6 & Max degree of $z$ in polynomials before GB. \\
      7 & Proportion of polynomials with $x$ before GB. \\
      8 & Proportion of polynomials with $y$ before GB. \\ 
      9 & Proportion of polynomials with $z$ before GB.\\ 
      10 & Proportion of monomials with $x$ before GB.\\
      11 & Proportion of monomials with $y$ before GB.\\ 
      12 & Proportion of monomials with $z$ before GB.\\
      13 & Number of polynomials after GB.\\
      14 & \texttt{TNoI} after GB.\\
      15 & \texttt{stds} after GB.\\
      16 & \texttt{tds} of polynomials after GB.\\
      17 & Max degree of $x$ in polynomials after GB.\\
      18 & Max degree of $y$ in polynomials after GB.\\
      19 & Max degree of $z$ in polynomials after GB.\\
      20 & Proportion of polynomials with $x$ after GB. \\
      21 & Proportion of polynomials with $y$ after GB. \\ 
      22 & Proportion of polynomials with $z$ after GB.\\ 
      23 & Proportion of monomials with $x$ after GB.\\
      24 & Proportion of monomials with $y$ after GB.\\ 
      25 & Proportion of monomials with $z$ after GB.\\
      26 & $\log_2$(\texttt{TNoI} before GB) - $\log_2$(\texttt{TNoI} after GB)\\
      27 & $\log_2$(\texttt{stds} before GB) - $\log_2$(\texttt{stds} after GB)\\
      28 & $\log_2$(\texttt{tds} before  GB) - $\log_2$(\texttt{tds} after  GB) \\ \hline
    \end{tabular}
\end{table}

As an example, consider sets of polynomials
\begin{align*}
E &:= \{
-12yz-3z,  \quad
17x^2-6,  \quad
-2yz+5x
\},
\\
F &:= \{
-2yz-9y,  \quad
-15x^2-19y,  \quad
6xz+3
\}.
\end{align*}
The GB computed for $E$ is
\[
G := \{
17x^2-6,  \quad
4y+1,  \quad
z+10x
\}.
\]
Then the \emph{all features} vector becomes
\begin{align*}
&\big[ 
\textstyle 12, 12, 2, 2, 1, 1, \frac{2}{3}, \frac{2}{3}, \frac{2}{3}, \frac{1}{3}, \frac{5}{12}, \frac{5}{12}, \\
& \qquad \textstyle 6, 10, 10, 2, 2, 1, 1, \frac{2}{3}, \frac{1}{2}, \frac{1}{2}, \frac{1}{3}, \frac{1}{3}, \frac{1}{4}, \\
& \qquad 0.263, 2.263, 0 
\big],
\end{align*} 
with the \emph{before features} and \emph{after features} vectors formed from the first and second lines respectively.

The feature generation process was applied to create the three training sets separately (although the feature labels used were all as in Table \ref{table:CSBAllFeatures}). 
Each problem was labelled $+1$ if GB preconditioning is beneficial for CAD construction, or $-1$ otherwise. 
After feature generation the training data was standardised so each feature had zero mean and unit variance across the training set.  The same standardisation was then applied to features in the test set.  

\subsection{Initial experiment and results}
\label{SUBSEC:CSBInitial}

The 1062 problems were partitioned into 80\% training (849 problems) and 20\% test (213 problems), stratified to maintain relative proportions of positive and negative examples. 

The classification was done in \textsc{SVM-Light} using the \emph{radial basis function} (RBF) kernel; with parameter values chosen to optimize MCC using a grid-search optimisation procedure with a five-fold stratified cross validation used.  The details are the same as for Case Study A so we refer to Section \ref{SUBSEC:CSAOptimization} for details.  This procedure was repeated for the three feature sets.

The classification accuracy was used to measure the efficacy of the machine learning  decisions.  The test set of 213 problems contained 159 positive samples and 54 negative samples (i.e. 75\% of the test problems benefited from GB preconditioning). 

The results of the machine learned choices are summarised in Table~\ref{tab:CSBnumof}.
First we note that when making a choice based on the \emph{before features} training set 75\% of the problems were predicted accurately. I.e. making a decision based on these features results in no more correct decisions than blindly deciding to GB precondition each and every time.  However, the other two feature sets resulted in superior decisions.  
Although only a small improvement on preconditioning blindly, the reader should recall that the wrong choice can give large changes to the size of the CAD or even change the tractability of the problem \cite{BH91, WBD12_GB}. 

The results certainly indicate that the features of the GB itself are required to decide whether to use the preconditioning.  They seem to indicate that the features from before the GB was applied actually hinder the learning process, but this would be an unsound conclusion.  Earlier research showed that a variable completely useless by itself can provide a significant performance improvement when taken in conjunction with others \cite{GE03}.  To be confident about which features were significant and which were superfluous, further feature selection experiments are required, as described in the next section.  We will see that the optimal feature subset must contain features from both before and after the GB computation.

\begin{table}[t]
  \caption{Accuracy of predictions before feature selection}
  \label{tab:CSBnumof}
\centering
   \def\arraystretch{1.2}%
    \begin{tabular}{lcc}
      \hline
      Feature Set 		 	& Number 	& \% of test set\\  \hline \hline
      All features 		 	& 162 		& 76\% \\
      Before features 	 	& 159 		& 75\% \\
      After features 	 	& 167 		& 78\% \\ \hline
    \end{tabular}
\end{table}

\subsection{Feature selection}
\label{SUBSEC:CSBSelection}

The feature selection experiments were conducted with \textsc{Weka} (Waikato Environment for Knowledge Analysis) \cite{HFHPRW09}, a Java machine learning library which supports tasks such as data preprocessing, clustering, classification, regression and feature selection. Each data point is represented as a fixed number of features. The inputs are samples of $29$ features, where the first $28$ are the real-valued features from Table \ref{table:CSBAllFeatures}, and the final one is a \emph{nominal} feature denoting its class. 

\subsubsection{The filter method}
\label{SUBSUBSEC:Filter}

First we applied a correlation based feature selection method as described in \cite{HH03}. Unlike other filter methods these measure the rank of feature subsets instead of individual features \cite{Hall2000}.  A feature subset which contains features highly correlated with the class but uncorrelated with each other is preferred.  

The metric below is used to measure the quality of a feature subset, and takes into account feature-class correlation as well as feature-feature correlation. 
\begin{equation}
\label{eq:Gs}
G_s = \frac{k\overline{r_{ci}}}{\sqrt{k+k(k-1)\overline{r_{ii^{'}}}}}
\end{equation}
Here, $k$ is the number of features in the subset, $\overline{r_{ci}}$ denotes the average feature-class correlation of feature $i$, and $\overline{r_{ii^{'}}}$ the average feature-feature correlation between feature $i$ and $i'$. 
The numerator of equation (\ref{eq:Gs}) indicates how much relevance there is between the class and a set of features, while the denominator measures the redundancy among the features. The higher $G_s$, the better the feature subset. 

To apply this heuristic we must calculate the correlations.
With the exception of the class attribute all 28 features are continuous, so in order to have a common measure for computing the correlations we first discretize using the method of Fayyad and Irani \cite{FI93}. After that, a correlation measure based on the information-theoretical concept of entropy is used: a measure of the uncertainty of a random variable.  We define the \emph{entropy of a variable} $X$  \cite{Shannon2001} as 
\begin{equation}
H(X)= \textstyle -\sum_{i} p(x_i)\log_2\big(p(x_i)\big).
\end{equation}
The entropy of $X$ after observing values of another variable $Y$ is then defined as
\begin{equation}
\textstyle H(X|Y) = -\sum_{j} p(y_j) \sum_{i} p(x_i|y_j) \log_2\big(p(x_i|y_j)\big),
\end{equation}
where $p(x_i)$ is the prior probabilities for all values of $X$, and $p(x_i|y_i)$ is the posterior probabilities of $X$ given the values of $Y$.
The \emph{information gain (IG)} measures the amount by which the entropy of $X$ decreases by additional information about $X$ provided by $Y$ \cite{Quinlan1986}.  It is given by
\begin{equation}
IG(X, Y) = H(X) - H(X|Y). 
\end{equation}
The \emph{symmetrical uncertainty (SU)} (a modified information gain measure) is then used to measure the correlation between two discrete variables (X and Y) \cite{PTVF92}: 
\begin{equation}
SU(X, Y)=2.0 \times \bigg(\frac{H(X) - H(X|Y)}{H(X)+H(Y)}\bigg).
\end{equation}

Treating each feature as well as the class as random variables, we can apply this as our correlation measure. More specifically, we simply use $SU(c, i)$ to measure the correlation between a feature $i$ and a class $c$, and $SU(i, i^{'})$ to measure the correlation between features $i$ and $i'$. These values are then substituted as $\overline{r_{ci}}$ and $\overline{r_{ii^{'}}}$ in equation (\ref{eq:Gs}). 

Recall that our aim here is to find the optimal subset of features which maximises the metric given in equation (\ref{eq:Gs}). 
The size of our feature set is 28 meaning there are
\[
2^{28}-1 \simeq 2.7 \times 10^8
\]
possible subsets, too many for exhaustive search. 
Instead a greedy stepwise forward selection search strategy was used for searching the space of feature subsets, which works by adding the current best feature at each round. The search begins with the empty set, and in each step the metric from equation (\ref{eq:Gs}) is computed for every single feature addition, and the feature with the best score improvement added.  If at some step none of the remaining features provide an improvement, the algorithm stops, and the current feature set is returned.  The best feature subset found with this method (which may not be the absolute optimal subset of features) is shown in Table \ref{tab:CSBfilter}, ordered by importance. 

\begin{table}[t]
  \caption{Feature selection by the filter method}
  \label{tab:CSBfilter}
\centering
\def\arraystretch{1.2}%
    \begin{tabular}{c l}
      \hline
      Feature Number & Description \\  \hline \hline
      14 	& \texttt{TNoI} after GB.\\
      13 	& Number of polynomials after GB.\\
      2 	& \texttt{stds} before GB.  \\
      26 	& $\log_2$(\texttt{TNoI} before GB) - $\log_2$(\texttt{TNoI} after GB)\\ 
      21 	& Proportion of polynomials with $y$ after GB.\\
      15 	& \texttt{stds} after GB.\\
      23 	& Proportion of monomials with $x$ after GB.\\
      19 	& Max degree of $z$ in polynomials after GB.\\
      25 	& Proportion of monomials with $z$ after GB.\\
      27 	& $\log_2$(\texttt{stds} before GB) - $\log_2$(\texttt{stds} after GB)\\ \hline
    \end{tabular}
\end{table}

\subsubsection{The wrapper method} 
\label{SUBSUBSEC:wrapper}

The wrapper feature selection method evaluates attributes using accuracy estimates provided by the target learning algorithm. Evaluation of each feature set was conducted with a SVM with RBF kernel function. The SVM algorithm is run on the dataset, with the same data partitions as used previously. Similarly, a five-fold cross validation was carried out. The feature subset with the highest average accuracy was chosen as the final set on which to run the SVM algorithm. 

In each training / validation fold starting with an empty set of features: each feature was added; a model was fitted to the training dataset; the classifier was then tested on the validation set. This was done on all the features giving a score for each reflecting the accuracy of the classifier. The final score for each feature was its average over the five folds. Having obtained a score for all features in the manner above, the feature with the highest score was then added in the feature set. The same greedy procedure as used for the filter method in Section \ref{SUBSUBSEC:Filter} was applied to obtain the best feature subset.   

Due to the large number of cases, the parameters $(C, \gamma)$ were selected from an optimised sub range instead of the full grid search used previously.  This suffices to demonstrate the performance of a reduced feature set.  The experiments of Section \ref{SUBSEC:CSBInitial} found (for all three feature sets) that $C$ taken from $\{2^{5}, 2^{6}, 2^{7}, 2^{8}, 2^{9}, 2^{10}\}$ and $\gamma$ taken from $\{2^{-5}$, $2^{-6}$, $2^{-7}, 2^{-8}, 2^{-9}, 2^{-10}\}$ provided good classifier performance.

The 36 pairs of $(C, \gamma)$ values were tested and an optimal feature subset with the highest accuracy was found for each. Then the one with the highest accuracy was selected as the final set, which is shown in Table \ref{tab:CSBWrapper} ordered by importance.  We see that most of the features selected (9, 12 and 22) related to variable $z$. 
Recall that the projection order used in the CAD was always $x \prec y \prec z$, i.e. the variable $z$ is projected first.  Hence it makes sense that this variable would have the greatest effect and thus be identified in the feature selection.

\begin{table}[t]
\caption{Feature selection by the wrapper method}
\label{tab:CSBWrapper}
\centering
    \begin{tabular}{c l}
      \hline
      Feature Number & Description \\  \hline \hline
      14 	& \texttt{TNoI} after GB.\\
      9 	& Proportion of polynomials with $z$ before GB.\\ 
      22 	& Proportion of polynomials with $z$ after GB. \\
      4 	& Max degree of $x$ in polynomials before GB. \\
      12 	& Proportion of monomials with $z$ before GB.\\ \hline
    \end{tabular}
\end{table}

\begin{figure}[t]
\centering
\includegraphics[width=0.40\textwidth]{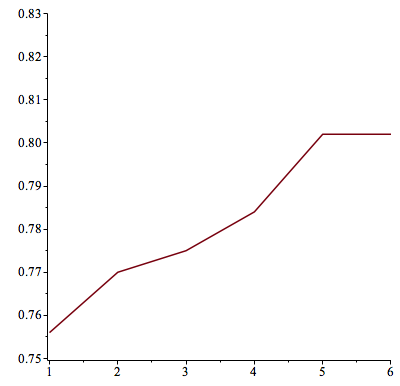}
  \caption{Performance of a sample classifier with different sizes of feature sets.  The vertical axis measures prediction accuracy and the horizontal the number of features.}
  \label{fig:CSBPerformance}
\end{figure}

We examined the performance on further reduced feature sets, obtained by the feature ranking of the wrapper method. Figure~\ref{fig:CSBPerformance} shows the overall prediction accuracies of a sample classifier. We see that when using a single feature (the best ranked feature was TNoI after GB for both filter and wrapper methods) this predictor achieved an accuracy score of 0.756 in this run.  Then the performance steadily increases with the size of the feature set until the fifth feature; but taking any sixth feature gives no further improvement and hence resulted in the cut-off chosen by the wrapper method.

\begin{figure}[t]
\begin{center}
\includegraphics[width=0.6\textwidth]{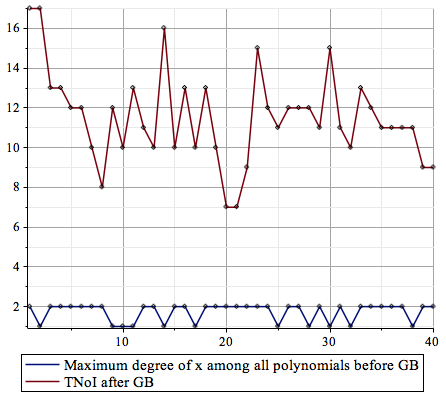}
  \caption{Values of features 4 and 14 for 40 misclassified data points}
  \label{fig:CSBerror1}
\end{center}
\end{figure}

\begin{figure}[t]
\begin{center}
\includegraphics[width=0.6\textwidth]{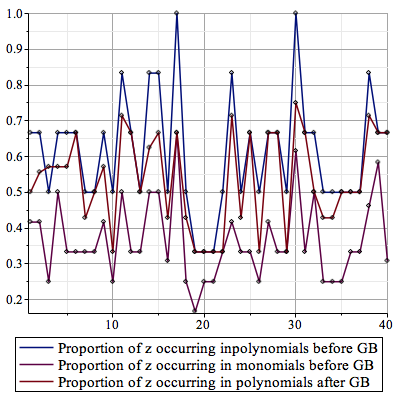}
  \caption{Values of features 9, 12 and 22 for 40 misclassified data points}
  \label{fig:CSBerror2}
\end{center}
\end{figure}

As the wrapper method identified only a few features an error analysis on the misclassified data points is feasible. Figure~\ref{fig:CSBerror1} shows 40 misclassified points and their values for features 4 and 14, while Figure~\ref{fig:CSBerror2} shows the remaining features (9, 12, and 22).
It is interesting that feature 4 of all misclassified samples is either 1 or 2, when for the whole dataset roughly a third of samples had this feature value 3 or 4.  This indicates that the algorithm performs better on instances with a higher maximum degree of $x$ among all polynomials before GB preconditioning.

\subsubsection{Results with reduced feature sets}
\label{SUBSUBSECT:FeatureResults}

Having obtained the reduced feature sets, we ran the experiment again to evaluate the new choices.  The dataset was again repartitioned into 80\% training and 20\% test set, stratified to maintain relative class proportions in both training and test partitions. Again, a five-fold cross validation and a finer grid-search optimisation procedure over the original range of ($C$, $\gamma$) pairs was conducted.
The choices with maximum averaged MCC 
were selected and the resulting classifier was then evaluated. 

The testing data was also reduced to contain only the features selected.  The classification accuracy was used to measure the performance of the classifier. 
In order to better estimate the generalisation performance of classifiers with reduced feature sets, the data was permuted and partitioned into 80\% training and 20\% test again and the whole process was repeated 50 times. For each run, each training set was standardised to have zero mean and unit variance, with the same offset and scaling applied subsequently to the corresponding test partition. 

\begin{figure}[t]
\begin{center}
\includegraphics[width=0.55\textwidth]{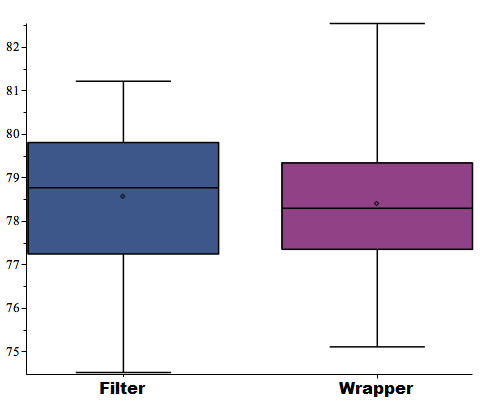}
  \caption{Boxplots of prediction accuracys from 50 runs of the 5-fold cross validation with the suggested feature sets from the two feature selection methods.}  \label{fig:CSBFeatureBox}
\end{center}
\end{figure}

Figure~\ref{fig:CSBFeatureBox} shows boxplots of the accuracies generated by 50 runs of the five-fold cross validation. Both reduced feature sets generated similar results and show an improvement on the base case where GB preconditioning is always used before CAD construction. 

The average overall prediction accuracy of the filter subset and the wrapper subset is 79\% and 78\% respectively (Figure~\ref{fig:CSBPerformance} shows a higher rate but that was just for one sample run).  All 50 runs of the wrapper subset performed above the base line, while the top three quartiles of the results of both sets achieve higher than 77\% percentage accuracy.   

\subsection{Comparison with the human developed heuristic}
\label{SUBSEC:TNoI}

We may compare the machine learned choice with the human developed \texttt{TNoI} heuristic \cite{WBD12_GB}, whose performance on the 213 test problems is shown in Table \ref{tab:TNoI}.  It correctly predicted whether GB preconditioning was beneficial for 118 examples, only 55\%.  So for this dataset it would have been better on average to precondition blindly than to make a decision using \texttt{TNoI} alone.  
The \texttt{TNoI} heuristic performed better in the experiments by Wilson et al. \cite{WBD12_GB}.  Those experiments involved only 22 problems (compared to 213 in the test set here) but they were human constructed to have certain geometric properties while the ones here were random.  

We also note that the \texttt{TNoI} heuristic actually performed quite differently for positive and negative examples of our dataset, as shown by the separated data in Table \ref{tab:TNoI}.  It was able to identify most of the cases where GB-preconditioning is detrimental but failed to identify many of the cases where it was beneficial.  The \texttt{TNoI} after GB was identified as important by both feature methods, but it seems to need to be used in conjunction with other features to be effective here.

\begin{table}[t]
  \caption{The performance of the \texttt{TNoI}-based heuristic \cite{WBD12_GB}}
  \label{tab:TNoI}
\centering
    \begin{tabular}{lccc}
      \hline
      					& Total & Correct Prediction  \\
      \hline \hline
      GB beneficial 	& 158   & 77  & (48\%) \\
      GB not beneficial & 54    & 41  & (76\%) \\
      Total             & 213   & 118 & (55\%) \\
      \hline
    \end{tabular}
\end{table}

\section{Final Thoughts}
\label{SEC:Conclusion}

\subsection{Summary}
\label{SUBSEC:Summary}

We have presented two case studies that apply machine learning techniques (specifically support vector machines) to make choices for CAD problem instances: the variable ordering to use and whether to precondition with a Gr\"obner Basis.   Both case studies showed the potential of machine learning to outperform human developed heuristics.

For Case Study A the experimental results confirmed our thesis, drawn from personal experience, that no one heuristic is superior for all problems and the correct choice will depend on the problem.  Each of the three heuristics tested had a substantial set of problems for which they were superior to the others.  Using machine learning to select the best CAD heuristic yielded considerably better results than choosing one heuristic at random, or just using any of the individual heuristics in isolation, indicating there is a relation between the simple algebraic features and the best heuristic choice.  This could lead to the development of a new individual heuristic in the future.  We note that we observed Brown's heuristic to be the most competitive for our example set, despite it involving less computation than the others.  This heuristic was presented during an ISSAC tutorial in 2004 \cite{Brown2004}, but does not seem to be formally published.  It certainly deserves to be better known.

For Case Study B a machine learned choice on whether to precondition was found to yield better results than either always preconditioning blindly, or using the previously human developed \texttt{TNoI} heuristic \cite{WBD12_GB}.  Two feature selection experiments showed that a small feature subset could be used to achieve this. The two subsets identified were different but both needed features from before and after the GB preconditioning.  For one, having fewer features actually improved the learning efficiency.   The smaller improvement achieved by machine learning in Case Study B is to be expected since it was a binary choice in which one option is known to usually be better.  It is still significant as making the wrong choice can greatly increase the size of the CAD, or even change the tractability of the problem. 

\subsection{Datasets}

As discussed in Section \ref{SUBSEC:Data} we had some difficulties obtaining suitable data for the experiments.  For the former experiment we had to employ a dataset from a different field (SMT solving) to obtain enough problems to train the classifier.  
This restricted the problems to those with a single block of existential quantifiers.  In the latter experiment we had to create a brand new dataset of random polynomials.  
We acknowledge that it would be preferable to run such experiments on an established dataset of meaningful problems but this was not simply not available.  We emphasise the interesting finding that the established \texttt{nlsat} dataset has a hidden uniformity despite its size\footnote{something that was noted in multiple presentations at the inaugural International Workshop on Satisfiability Checking and Symbolic Computation in 2016 and is now being addressed by the \textsf{SC}$^2$ Project \cite{AAB+16a}.}.  It should be noted that the randomly created dataset used matched the previously reported results on the topic of investigation \cite{BH91, WBD12_GB}.

The experiments described involved building 1000s of CADs, certainly the largest such experiment that the authors are aware of.  For comparison, the best known previous study on CAD variable ordering heuristics \cite{DSS04} tested with six examples.  

\subsection{Future work}
\label{SUBSEC:Future}

First, as noted above it is desirable to re-run both experiments on further datasets of problems as they become available.  There is a large set derived from university mathematics entrance exams \cite{KIMA16}, which but may be publicly available in the future.
An obvious extension of Case Study A would be to test additional heuristics, such as the greedy sotd heuristic \cite{DSS04}, one based on the number of full dimensional cells \cite{WEBD14}, or those developed for CAD by Regular Chains \cite{EBDW14}.  For Case Study B there are now further CAD optimisations for problems with multiple equalities under development \cite{EBD15, ED16a, DE16} which may affect the role of GB preconditioning from CAD.  Also, as pointed out by one of the referees, one could consider trying to classify whether to reduce individual inequalities with the GB.

We choose SVMs with an RBF kernel for these experiments based on some similar work for theorem provers but it would be interesting to see if other machine learning methods could offer similar or even better selections.  Further improvements may also come from more work on the feature selection.  The features used here were all derived from the polynomials involved in the input.  One possible extension would be to consider also the type of relations present and how they are connected logically.  This is likely to be increasingly beneficial as new CAD theory is developed which takes this into account \cite{BDEMW13, BDEMW16, EBD15, BCDEMW14, JdM12, Brown2015}.  

These two experiments were conducted independently: the possibility of GB preconditioning was not considered in Case Study A while in Case Study B the variable ordering for CAD and the monomial ordering for GB were fixed.   In reality such decisions may need to be taken in tandem and it is possible that preconditioning could change the best choice of variable ordering and the variable ordering change whether we would precondition.  How best to tackle this, other than applying heuristics in serial, is another topic for future consideration.


Finally, we note that CAD is not unique amongst computer algebra algorithms in requiring the user to make such a choice of problem formulation.  Computer algebra systems (CASs) often have a choice of possible algorithms to use when solving a problem.  Since a single formulation or algorithm is rarely the best for the entire problem space, CASs usually use \emph{meta-algorithms} to make such choices, where decisions are based on some numerical parameters \cite{Carette2004}.  These are often not as well documented as the base algorithms, and may be rather primitive.  
To the best of our knowledge, the present paper appears to be the first applying machine learning to problem formulation for computer algebra. The positive results should encourage investigation of similar applications throughout the field.

\section*{Acknowledgements}
This work was supported by EPSRC grant EP/J003247/1; the European Union's Horizon 2020 research and innovation programme under grant agreement No 712689 (\textsf{SC}$^2$); and the China Scholarship Council (CSC). 
The authors acknowledge both the anonymous referees of the present paper, and those of \cite{HEWDPB14} and \cite{HEDP16}, whose comments also helped improve this article.

\bibliographystyle{plain}
\bibliography{CAD}

\end{document}